\documentclass[12pt]{article}
\usepackage{amsfonts,amssymb,amsmath,epsfig,graphicx}
\usepackage{theorem,calc}
\usepackage{pgf,tikz}
\usepackage{hyperref}
\numberwithin{equation}{section}

\newcommand{\nonu}{\nonumber \\}

\newcommand{\mb}[1]{\quad\mbox{#1}\quad}
\newcommand{\beq}{\begin{equation}}
\newcommand{\eeq}{\end{equation}}
\newcommand{\bea}{\begin{eqnarray}}
\newcommand{\eea}{\end{eqnarray}}
\newcommand{\beano}{\begin{eqnarray*}}
\newcommand{\eeano}{\end{eqnarray*}}

\newcommand{\eps}{\epsilon}

\DeclareMathOperator{\res}{Res}

      \def\cB{{\cal B}}

\def\fm{{\mathfrak m}}
\def\fn{{\mathfrak n}}

\newcommand{\wh}[1]{\widehat{#1}}

\newcommand{\half}{\frac{1}{2}}

\newcommand{\ket}[1]{| #1 \rangle}

\newcounter{heure}
\newcounter{minute}

\usepackage{color}

\begin{document}

\renewcommand{\thefootnote}{\arabic{footnote}}
\setcounter{footnote}{0}
\newpage
\setcounter{page}{0}

\pagestyle{empty}

\null
\vfill
\begin{center}

\begin{minipage}{15cm}
\begin{center}
{\Huge \textsf{Eigenvectors of open XXZ and ASEP models
for a class of non-diagonal boundary conditions}}
\end{center}
\null

\vfill

\vspace{1cm}

{\large  \textbf{N. Cramp{\'e}}$^{a}$, \textbf{E. Ragoucy}$^{b}$ and 
\textbf{D. Simon}$^{c}$ }
\vspace{10mm}

\emph{$^a$ LPTA, CNRS and Universit\'e Montpellier II,}\\

\emph{Place Eug{\`e}ne Bataillon,
34095 Montpellier Cedex 5, France}
\\
E-mail: ncrampe@um2.fr
\vspace{1cm}

\emph{$^b$ LAPTH, CNRS and Universit\'e de Savoie}\\
\emph{9 chemin de Bellevue, BP 110, 74941, Annecy-Le-Vieux Cedex, 
France}
\\
E-mail: ragoucy@lapp.in2p3.fr
\vspace{1cm}

\emph{$^c$ LPMA, Universit\'e Pierre et Marie Curie,}

\emph{Case Courrier 188, 4 place Jussieu,
75252 Paris Cedex 05, France}\\
E-mail: damien.simon@upmc.fr

\vspace{1cm}

\begin{abstract}
We present a generalization of the coordinate Bethe ansatz that 
allows us
to solve integrable open XXZ and ASEP models with non-diagonal 
boundary 
matrices, provided their parameters obey some relations. These 
relations extend the ones already known in the literature in the 
context of algebraic or functional Bethe ansatz. 
The eigenvectors are represented as sums over cosets of the 
$BC_n$ Weyl group. 
\end{abstract}
\end{minipage}
\end{center}

\vfill
\vfill
\rightline{LAPTH-040/2010}

\baselineskip=16pt

\newpage
\pagestyle{plain}
\section{Introduction}

In the framework of integrable systems, the study of open spin chains 
with integrable boundaries have been developed a long time ago 
\cite{cherednik,sklyanin}. There, it has been shown that the model is 
integrable provided the two matrices characterizing the boundaries 
obey some algebraic relations (the so-called reflection equation). 
However, although the integrability has been proven, the explicit 
resolution 
(eigenvalues and eigenvectors of the Hamiltonian)
of the models is not known in its full generality. In fact, for a 
long 
time, only the case of diagonal boundary matrices was solved, for 
different kind of models, e.g. open XXX \cite{Gau} and its $su(N)$ 
\cite{byebye} or $su(N\vert M)$ \cite{RS,zaka}
generalizations, open XXZ \cite{alc} and its generalizations 
\cite{mnsymm,ACDFR2,done2,sam},
using different versions of the 
Bethe ansatz (analytical, algebraic or functional).
However, the classifications of boundary matrices (obeying a 
reflection equation) \cite{dvgr2, gand, selene, nondiag-sol} clearly shows that 
non-diagonal solutions do 
exist, although explicit solutions for the eigenvalue problem were 
not known. The problem laid 
essentially in the construction of a reference state (a particular 
Hamiltonian eigenvector) that allows to initiate the procedure. 
Indeed, when the boundary matrices were not diagonal (or at least not
simultaneously diagonalizable), the existence of this reference state 
was not ensured.

Recently, different approaches have been developed to overcome 
these difficulties, such as gauge transformations that allow to go to a 
diagonal basis \cite{Cao}, or fusion 
relations for TQ relations that do not need the existence of a 
reference state \cite{nepo,tq}. In all these
cases, the boundary matrices can be non-diagonal, but the 
parameters entering their definition need to satisfy some constraints.
Note also the original approach \cite{BK} that uses another 
presentation of the 
reflection algebra, called $q$-Dolan-Grady relation, as well as the 
`generalized' functional ansatz developed in \cite{galleas}. Both avoid the 
use of constraints.

The Asymmetric Simple Exclusion Process (ASEP) is an 
out-of-equilibrium statistical physics representation of the 
Temperley-Lieb algebra \cite{simon09}, on which the XXZ Hamiltonian 
is based. Many exact probabilistic results 
\cite{dehp,schuetz1,schuetz2,jafarpour}, that are not necessarily 
based on the integrability of the model, have been obtained in the 
past fifteen years and shed some new light on integrable results. The 
ASEP notations, as described in \cite{dGE}, are presented in this 
paper and a specific section \ref{sec:matrixansatz} is dedicated to 
the comparison of our results with old results on the ASEP, such as 
the matrix ansatz \cite{dehp,mallicksandow,esr,corteel}.

In this paper, we present a construction based on the 
coordinate Bethe ansatz \cite{bethe} for open XXZ and ASEP models, allowing the use of 
non-diagonal boundary 
matrices. As for the other approaches, the boundary matrices need to 
obey some constraints, but the ones we find are more general than 
those already known.

We present now the structure of the paper: section 
\ref{sec:defandnotations} defines the models and introduces the 
notations needed for the next sections; section \ref{sec:basis} 
describes the different sets of reference states and the sets of 
exceptional points they lead to. Section \ref{sec:CBA} gives the 
structure of the coordinate Bethe ansatz as a combination of the 
reference states and explains the role of the $BC_n$ Weyl group and gives in details the Bethe equations that correspond to the two sets of specific points. Finally, we conclude on two open problems: section \ref{sec:matrixansatz} 
presents a short discussion of the relation between the specific points and 
other previous results about the exclusion process such as the matrix ansatz; a discussion on the completeness of the spectrum and of the eigenvectors is tackled in section \ref{subsec:completeness}.

\section{XXZ and ASEP models with non-diagonal boundaries}
\label{sec:defandnotations}

The Markov transition matrix for the open ASEP model is given by
\beq \label{eq:hamasep}
W=\widehat K_1+K_L+\sum_{j=1}^{L-1}w_{j,j+1}\,,
\eeq
where the indices indicate the spaces in which the following matrices 
act non trivially
\bea
w=\left(
\begin{array}{c c c c}
 0 & 0 & 0 & 0\\
 0 &-q & p & 0\\
 0 & q &-p & 0\\
 0 & 0 & 0 & 0
\end{array}
\right)
\mb{,}
\widehat K=\left(
\begin{array}{c c}
 -\alpha & \gamma e^{-s} \\
 \alpha e^{s} & -\gamma 
\end{array}
\right)
\mb{and}
K=\left(
\begin{array}{c c}
 -\delta & \beta  \\
 \delta  & -\beta 
\end{array}
\right)\,.
\eea

It is well-established \cite{ssa,esr,dGE} that this ASEP model is 
related by a 
similarity transformation to 
the following integrable open XXZ model 
\bea \label{eq:hamxxz}
H=\wh B_1 + B_L -\half\sum_{j=1}^{L-1}\big(\sigma^x_j\sigma^x_{j+1}
+\sigma^y_j\sigma^y_{j+1}-\cos\eta~ \sigma^z_j\sigma^z_{j+1}\big)\,,
\eea
where $\sigma$ are the usual Pauli matrices and
\bea
\widehat B&=&
\frac{\sin\eta}{\cos\omega_-+\cos\delta_-}
\left(
\begin{array}{c c}
\frac i2 (\cos\omega_--\cos\delta_-)-\sin\omega_-  & e^{-i\theta_1}  
\\
 e^{i\theta_1}  & -\frac i2 (\cos\omega_--\cos\delta_-)-\sin\omega_-
\end{array}
\right)
\,,
\\
 B&=&
\frac{\sin\eta}{\cos\omega_++\cos\delta_+}
\left(
\begin{array}{c c}
-\frac i2 (\cos\omega_+-\cos\delta_+)-\sin\omega_+  & e^{-i\theta_2}  
\\
 e^{i\theta_2}  & \frac i2 (\cos\omega_+-\cos\delta_+)-\sin\omega_+
\end{array}
\right)
\,.
\eea
The explicit form of the transformation can be found in e.g. 
\cite{dGE}. 
We reproduce it here for completeness:
\bea
W &=& -\sqrt{pq}\,U^{-1}\,H\,U
\mb{with} U=\otimes_{j=1}^L\left(\begin{array}{cc} 1 & 0 \\
0 & \xi\left(\sqrt{\frac{q}{p}}\right)^{j-1}\end{array}\right)
\,,\\
\sqrt{\frac{\alpha}{\gamma}} &=&-ie^{i\omega_{-}} \mb{,} 
\sqrt{\frac{\beta}{\delta}} =-ie^{i\omega_{+}}
\mb{,} \sqrt{\frac pq} = -e^{i\eta}\\
\xi\sqrt{\frac{\alpha}{\gamma}}\,e^s &=& e^{i\theta_{1}} \mb{and} 
\xi\sqrt{\frac{\delta}{\beta}}\left(\sqrt{\frac{q}{p}}\right)^{L-1} = 
e^{i\theta_{2}}\,,
\label{eq:asepXXZ}
\eea
where $\xi$ is an arbitrary (gauge) parameter that disappears in all 
the following computations.

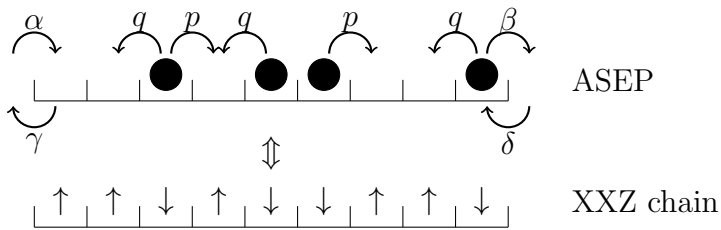
\begin{figure}
\begin{center}
 \begin{tikzpicture}[scale=0.7]
\draw (0,0) -- (9,0) ;
\foreach \i in {0,1,...,9}
{\draw (\i,0) -- (\i,0.4) ;}
\draw[->,thick] (-0.4,0.9) arc (180:0:0.4) ; \node at (0.,1.5) 
{$\alpha$};
\draw[->,thick] (0.4,-0.1) arc (0:-180:0.4) ; \node at (0.,-0.8) 
{$\gamma$};
\draw  (2.5,0.5) circle (0.3) [fill,circle] {};
\draw  (4.5,0.5) circle (0.3) [fill,circle] {};
\draw  (5.5,0.5) circle (0.3) [fill,circle] {};
\draw  (8.5,0.5) circle (0.3) [fill,circle] {};
\draw[->,thick] (2.4,0.9) arc (0:180:0.4); \node at (2.,1.5) {$q$};
\draw[->,thick] (2.6,0.9) arc (180:0:0.4); \node at (3.,1.5){$p$};
\draw[->,thick] (4.4,0.9) arc (0:180:0.4); \node at (4.,1.5){$q$};
\draw[->,thick] (5.6,0.9) arc (180:0:0.4); \node at (6.,1.5) {$p$};
\draw[->,thick] (8.4,0.9) arc (0:180:0.4); \node at (8.,1.5){$q$};
\draw[->,thick] (8.6,0.9) arc (180:0:0.4) ; \node at (9.,1.5) 
{$\beta$};
\draw[->,thick] (9.4,-0.1) arc (0:-180:0.4) ; \node at (9.,-0.8)
{$\delta$};
\node at (10.,0.4) [anchor=west] {ASEP};
\node at (4.5,-1.) {$\Updownarrow$};
\draw (0,-2.4) -- (9,-2.4) ;
\foreach \i in {0,1,...,9}
{\draw (\i,-2.4) -- (\i,-2.) ;}
\node at (2.5,-1.9) {$\downarrow$};
\node at (4.5,-1.9) {$\downarrow$};
\node at (5.5,-1.9) {$\downarrow$};
\node at (8.5,-1.9) {$\downarrow$};
\node at (0.5,-1.9) {$\uparrow$};
\node at (1.5,-1.9) {$\uparrow$};
\node at (3.5,-1.9) {$\uparrow$};
\node at (6.5,-1.9) {$\uparrow$};
\node at (7.5,-1.9) {$\uparrow$};
\node at (10.,-1.9) [anchor=west] {XXZ chain};
 \end{tikzpicture}
 \end{center}
 \caption{Asymmetric exclusion process for a system of size $L$ with 
two boundaries and mapping to the XXZ spin chain.}
 \label{fig:schemaasep}
\end{figure}

Throughout the paper we will stick to this ASEP notation, keeping in 
mind that by this similarity transformation we treat also the XXZ 
model 
with non-diagonal boundaries. However, in the ASEP model, all the 
parameters must be 
positive (transition rates) while there is not such constraint in the 
XXZ spin chain. 
To be as general as possible, we will not assume 
this constraint in the paper.

The vectors of the canonical basis used for the previous definitions 
are indexed by the spin value $\uparrow$ or $\downarrow$ for the 
XXZ-spin chain and corresponds to the number of particle 
$\tau_i\in\{0,1\}$ of each site in the ASEP (see figure 
\ref{fig:schemaasep}).

\section{Basis vectors for the coordinate Bethe 
ansatz\label{sec:basis}}

For a chain of $L$ sites with periodic or diagonal boundary 
conditions, it is easy to find 
one eigenstate, called pseudo-vacuum. It is usually chosen as $L$ 
spins up (resp. $L$ empty sites for the ASEP).
Then, excited states are constructed by adding some excitations 
that flip a given number $n$ of spins (resp. add $n$ particles in 
the ASEP).
The conventional Bethe methods (coordinate, algebraic or analytical) 
allow us 
to compute the linear combinations between the 
$\left(\begin{array}{c}L\\n\end{array}\right)$ 
excited states that diagonalize the Hamiltonian.

For an open chain with non-diagonal boundary (which is the case we 
want to 
deal with, see equations
(\ref{eq:hamasep}) and (\ref{eq:hamxxz})),
the first problem consists in finding the pseudo-vacuum. 
Different approaches have been elaborated to overcome this problem
as explained in the introduction. However, the coordinate Bethe 
ansatz, which is the 
historical method introduced by H. Bethe \cite{bethe}, has never been 
successfully applied.
To treat this problem, we must generalize the ansatz. A first step in 
this direction have been done in \cite{simon09}. The general 
strategy we adopt here can be summarized as follows:
\begin{itemize}
\item we do not choose 
anymore all spins up for the vacuum and spins down for the 
excitations. We take general vectors (see their explicit 
construction below),
\item Hamiltonian eigenfunctions will be constructed as linear 
combinations 
between states with $n$ excitations 
\textit{together with} $m$ $(<n)$ excitations,
\item the explicit forms of the excitations and of the vacuum depend 
on 
the total number of excitations in the state we consider.
\end{itemize}
More physically, one allows that excitations may be destroyed
or created by the boundaries with the restriction that
one cannot create more than $n$ excitations.
Without any loss of generality, we can always choose that one of 
the boundaries preserves the 
number of excitations. In this paper, we choose the right-hand-side one.

Finally, we found two different ways to fulfill these conditions. In 
both cases, some constraints between 
the parameters of the model appear (as it is already the case in the 
 approaches \cite{Cao,dGE} to non-diagonal boundaries). We will show 
that we 
can solve the problem (within the framework of coordinate Bethe 
ansatz) for  all the sets of constraints that have already been
produced in the 
literature as well as for some new sets.

The following subsection introduces some notations that are standard 
for the ASEP and that will be used throughout the paper. Then, in the 
next two subsections \ref{sec:first} and \ref{sec:second}, we will 
present the two 
different sets of vectors with the associated constraints.
Finally, in section \ref{sec:CBA}, we present their particular linear 
combinations diagonalizing the Hamiltonian.

\subsection{Boundary operators and a duality}
\label{sec:boundarynotations}

Two particular sets of 2-dimensional vectors are relevant in the 
study of the boudary dynamics. The first set corresponds to 
vectors that diagonalize the boundary operators. We choose the 
following arbitrary normalizations, that will become clearer later:
\begin{align}
\widehat{K} \begin{pmatrix} 1 \\ e^s/c_1 \end{pmatrix} &= \lambda_1 
\begin{pmatrix} 1 \\ e^s /c_1 \end{pmatrix}\,,\qquad
K \begin{pmatrix} 1 \\ c_L \end{pmatrix} = \lambda_L 
\begin{pmatrix} 1 \\ c_L \end{pmatrix} \,.
\end{align} 
There are two solutions for the first equation (resp. the second) 
given by $c_1=c_\pm(\alpha,\gamma)$ and $\lambda_1 = 
\lambda_\pm(\alpha,\gamma)$ (resp. $c_L=c_\pm(\beta,\delta)$ and 
$\lambda_L=\lambda_\pm(\beta,\delta)$) where  
$c_\pm(u,v)$ and $\lambda_\pm(u,v)$ are the  
roots of the two functions:
\begin{align}
P_{u,v}(X) &= uX +(u-v) -v/X  \mb{i.e.} c_+(u,v) = v/u \mbox{ and } 
c_-(u,v)=-1\,,\\
Q_{u,v}(X) &= X^2 + X(u+v) \mb{i.e.} \lambda_+(u,v)=0 \mbox{ and } 
\lambda_-(u,v)=-u-v\,.
\end{align}
These notations allow for a 
convenient way of parametrizing both boundaries at the same time.

The second set of relevant vectors satisfies the diagonal relations:
\begin{align}\label{eq:Kstar:def}
\left[ \widehat{K} - \begin{pmatrix} q & 0 \\ 0 & p 
\end{pmatrix}\right]\begin{pmatrix} 1 \\ e^s/c^*_1 \end{pmatrix}  = 
\lambda^*_1 \begin{pmatrix} 1 \\ e^s/c^*_1 \end{pmatrix}
\,, \quad
\left[ K + \begin{pmatrix} q & 0 \\ 0 & p 
\end{pmatrix}\right]\begin{pmatrix} 1 \\ c^*_L \end{pmatrix}  = 
(\lambda^*_L+p+q) \begin{pmatrix} 1 \\ c^*_L \end{pmatrix} 
\,.
\end{align}
As previously, the coefficients $c^*_1$ and $\lambda^*_1$ (resp. 
$c^*_L$ and $\lambda^*_L$) take the two possible values 
$c^*_\pm(\alpha,\gamma)$ and $\lambda^*_{\pm}(\alpha,\gamma)$ (resp. 
$c^*_\pm(\beta,\delta)$ and $\lambda^*_{\pm}(\beta,\delta)$), which 
are the respective zeroes of
\begin{align}
P_{u,v}^*(X) &= uX + (u-v+q-p) - v/X= P_{u,v}(X)+(q-p) \,,\\
Q_{u,v}^*(X) &= X^2+ X(u+v+p+q) + (qp+pu+qv)\,.
\end{align}
The explicit  values of $c^*_\pm(u,v)$ 
\footnote{These functions are sometimes called $\kappa_\pm(u,v)$. 
We change this notation to be consistent throughout the paper.} and 
$\lambda^*_\pm(u,v)$ are 
given by:
\begin{align}
c^*_\pm(u,v) &= \frac{p-q+v-u \pm 
\sqrt{(p-q+v-u)^2+4uv}}{2u} 
\,,\\
\lambda^*_\pm(u,v) &= \frac{-p-q-v-u \pm \sqrt{(p-q+v-u)^2+4uv}}{2}
\,.
\end{align}
The definitions \eqref{eq:Kstar:def} suggest the definition of the 
new operators:
\begin{subequations}
\label{eq:staroperator:def}
\begin{align}
\widehat{K}^* &= \widehat{K} - \begin{pmatrix} q & 0 \\ 0 & p 
\end{pmatrix}
\,,\qquad
K^* = K + \begin{pmatrix} q & 0 \\ 0 & p \end{pmatrix} 
\,, \\
w^* &=  w + \begin{pmatrix} q & 0 \\ 0 & p 
\end{pmatrix} \otimes I - I \otimes \begin{pmatrix} q & 0 
\\ 0 & p \end{pmatrix}  = \begin{pmatrix} 0 & 0 & 0 & 0 \\ 0 & 
-p & p & 0 \\ 0 & q & -q & 0 \\ 0 & 0 & 0 & 0 \end{pmatrix}
\end{align}
\end{subequations}
where $I$ is the two by two identity matrix.
One verifies that the matrix 
$W$ given in (\ref{eq:hamasep}) has the following second 
representation:
\begin{equation}
\label{eq:secondrepres}
W =\widehat{K}_1^* + K_L^* + \sum_{i=1}^{L-1} w_{i,i+1}^*\,.
\end{equation}
This representation will be essential in the definitions of the new 
exceptional points.

\subsection{First choice\label{sec:first}}

We first consider the matrix $W$ as written in eq. 
(\ref{eq:hamasep}). 
Let us define the family of vectors:
\begin{align}
\ket{\omega(u)}_i &= \begin{pmatrix} 1 \\ u (p/q)^{i-1} \end{pmatrix} 
\,,
\\
\ket{V}_i &= \begin{pmatrix} q-p \\ 0 \end{pmatrix}  
\,.
\end{align}
where $u$ is still an arbitrary parameter. 
{From} these elementary vectors, we build the tensor product over the 
sites $i$ to $j$ and write it as
\begin{equation}
\label{eq:def:Omega}
\ket{\Omega(u)}_i^j= 
\ket{\omega(u)}_i\ket{\omega(u)}_{i+1}\ldots\ket{\omega(u)}_j
\,.
\end{equation}
We now fix an integer $n$ and introduce the state with $n-m$ 
excitations at the ordered positions $1\leq x_{m+1}< \ldots < x_n 
\leq L$ defined as the $(\mathbb{C}^2)^{\otimes L}$-vector:
\bea
&&\ket{x_{m+1},\ldots,x_n} \ =\ 
\left(\sqrt{\frac{q}{p}}\right)^{(x_{m+1}-1)+\ldots+(x_n-1)} \times
\label{eq:def:state}\\
&&\quad\times\ 
\ket{\Omega(u_{m+1})}_1^{x_{m+1}-1}\,\ket{\omega(v_{m+1})}_{x_{m+1}}\,
\ket{\Omega(u_{m+2})}_{x_{m+1}+1}^{x_{m+2}-1}\,\ket{\omega(v_{m+2})}_{x_{m+2}}
\ \ldots\ \ket{\omega(v_n)}_{x_n}\,\ket{\Omega(u_{n+1})}_{x_n+1}^L\,.
\nonumber
\eea
The overall factor $(\sqrt{\frac{q}{p}})$ is introduced only in order 
to normalize the Bethe roots. The coefficients $u_m$ and $v_m$ are 
related through the recursion relation:
\begin{equation}
\label{eq:recursion:ui:vi}
 u_{m+1} = \frac{q}{p} u_m, \quad v_{m+1} =  \frac{q}{p} v_m\,,
\end{equation}
and the initial coefficients $u_1$ and $v_1$ are still arbitrary. 
These vectors correspond to states where $m$ excitations have left 
the 
system (through the left boundary). To clarify the notation, the state
with no excitation corresponds to $m=n$ and is given by
\begin{equation}
\ket{\emptyset}= \ket{\Omega(u_{n+1})}_{1}^{L}\,,
\end{equation}
while the state with one excitation ($m=n-1$) reads
\begin{equation}
\ket{x_{n}}= \left(\sqrt{\frac{q}{p}}\right)^{x_n-1}
\ket{\Omega(u_{n})}_1^{x_{n}-1}\,\ket{\omega(v_n)}_{x_n}\,\ket{\Omega(u_{n+1})}_{x_n+1}^L\,.
\end{equation}

The above states are product states and are linearly independent. We 
denote by $\cB_n$ the set of these  $\displaystyle \sum_{k=0}^n 
\left(\begin{array}{c} L\\k\end{array}\right)$ independent vectors.
We also need the vectors $|x_{m+1},\dots,\overline 
x_\alpha,\dots,x_n\rangle$ 
deduced from $|x_{m+1},\dots,x_n\rangle$ by replacing the vector 
$\ket{\omega(u)}$ in
 position $x_\alpha$ by $\ket{V}_{x_\alpha}$.

These vectors have been chosen such that we get for the Hamiltonian 
bulk part:
\begin{equation}
 w_{\beta,\beta+1}|x_{m+1},..,x_n\rangle=
\begin{cases}
 0 &\text{if } \beta,\beta+1\neq x_{m+1},\dots,x_n
\,;\\
\sqrt{pq}|..,x_j-1,..\rangle
-q|..,x_j,..\rangle +|..,\overline x_j,..\rangle
&\text{if } \beta+1=x_j\ ,\,\beta\neq x_{j-1}
\,;\\
\sqrt{pq}|..,x_j+1,..\rangle
-p|..,x_j,..\rangle -|..,\overline x_j,..\rangle
&\text{if } \beta=x_j\ ,\,\beta+1\neq x_{j+1}
\,;\\
|..,x_j,\overline x_{j+1},..\rangle
-|..,\overline x_j,x_{j+1},..\rangle
&\text{if } \beta=x_j\ ,\,\beta+1=x_{j+1}\,.
\end{cases}
\end{equation}
Remark the additional states appearing on the right hand side 
involving vector $|V\rangle$. 
Because of the alternating signs, they cancel each other in the 
Hamiltonian bulk part, 
but the first and last terms (boundary terms). It is interesting to 
remark that the same "telescopic" trick is also used in the proof of 
the matrix ansatz in section \ref{sec:matrixansatz} and in the 
definitions \eqref{eq:staroperator:def}.

The first and last terms $-\ket{V}_1$ and $\ket{V}_L$ of the 
telescopic sum must be absorbed by the boundary operators and this 
justifies the study of the operator \eqref{eq:Kstar:def}.

We now require that the left boundary diagonalizes 
$\ket{\omega(u_1)}_1$ and that the right boundary diagonalizes 
$\ket{\omega(u_{n+1})}_L$, hence  excitations are created neither on 
the right, nor on the left, when $n$ 
particles are already in the bulk. Thus, the action of the boundary 
operators must be
\begin{align}\label{eq:K1}
 \widehat K_1|x_{m+1},\dots,x_n\rangle &=
\begin{cases}
 \Lambda_1^{(m)}|x_{m+1},\dots,x_n\rangle 
+C_1^{(m)}|1,x_{m+1},\dots,x_n\rangle &\text{if } x_{m+1}>1\,,\\
\widetilde\Lambda_1^{(m)}|1,x_{m+2}\dots,x_n\rangle 
+D_1^{(m)}|x_{m+2},\dots,x_n\rangle + 
\ket{\overline{1},x_{m+2},..,x_n} &\text{if } x_{m+1}=1\,,
\end{cases} \\
 K_L|x_{m+1},\dots,x_n\rangle &=
\begin{cases}
 \Lambda_L|x_{m+1},\dots,x_n\rangle &\text{if } x_{n}<L\,,\\
\widetilde\Lambda_L|x_{m+1},\dots,x_{n-1}, L\rangle
-|x_{m+1},\dots,x_{n-1},\overline L\rangle&\text{if } x_{n}=L\,,
\end{cases} \label{eq:KL}
\end{align}
with the constraint
\begin{equation}
C_1^{(0)}=0 \label{eq:C1null}\,.
\end{equation} 
We introduce also the more compact form $\Lambda_1=\Lambda_1^{(0)}$ 
since this value will appear in the energy.

The constraints \eqref{eq:C1null} and \eqref{eq:KL} lead to the 
following trivial identifications with the parameters introduced in 
section \ref{sec:boundarynotations}:
\begin{align}
u_1 &= e^s / c_\epsilon(\alpha,\gamma)\,, &\Lambda_1 &= 
\lambda_\epsilon(\alpha,\gamma) \,,\\
(p/q)^{L-1} u_{n+1} &= c_{\epsilon'}(\beta,\delta)\,,  &\Lambda_L 
&= \lambda_{\epsilon'}(\beta,\delta) \,,\\
(p/q)^{L-1} v_n &= c^*_{\epsilon''}(\beta,\delta)\,, 
&\widetilde{\Lambda}_L &= \lambda^*_{\epsilon''}(\beta,\delta)+q\,,
\end{align}
where $\epsilon$, $\epsilon'$, $\epsilon'' \in \{+,-\}$ are 
arbitrary. The compatibility relation of the previous equalities with 
the recursion relations \eqref{eq:recursion:ui:vi} lead to four 
possible solutions summarized in Table~\ref{tab:cons1}. They 
correspond to the different choices of the signs $\epsilon$, 
$\epsilon'$ in the relation:
\begin{equation}
\label{eq:constraint1}
c_\epsilon(\alpha,\gamma) c_{\epsilon'}(\beta,\delta) = e^s 
\left(\frac{p}{q}\right)^{L-1-n}\,.
\end{equation}
The third sign $\epsilon''$ does not appear in this condition, nor in 
the energy, nor in the Bethe equations: it just corresponds to the 
full 
reflection of the excitations on the right boundary\footnote{Of 
course, 
we could have 
also chosen a full reflection on the left boundary instead.}.

\begin{table}[h b t]
\begin{centering}
\begin{tabular}{|c | c  || c | c || c |}
 \hline
$\Lambda_1$ & $\Lambda_L$  & $c_\epsilon(\alpha,\gamma)$ 
&$c_{\epsilon'}(\beta,\delta)$
& Constraints 
\\  
\hline
$0$         &  $0$        & $c_+(\alpha,\gamma) $  & $ 
c_+(\beta,\delta)$ & 
$ 
\frac{\alpha\beta}{\gamma\delta}e^s\left(\frac{p}{q}\right)^{L-1-n}=1$\\

\hline
$-\alpha-\gamma$         &  $-\beta-\delta$& $c_-(\alpha,\gamma) $ & 
$ c_-(\beta,\delta)$        
& $ e^s\left(\frac{p}{q}\right)^{L-1-n} = 1 $                 \\  
\hline
$-\alpha-\gamma$         &  $0$  & $ c_-(\alpha,\gamma)$ &  $ 
c_+(\beta,\delta)$    
& $ -\frac{\beta}{\delta}e^s\left(\frac{p}{q}\right)^{L-1-n} = 
1$                 \\  
\hline
$0$         &  $-\beta-\delta$ & $ c_+(\alpha,\gamma)$ & $  
c_-(\beta,\delta)$
 &  $ - \frac{\alpha}{\gamma}e^s\left(\frac{p}{q}\right)^{L-1-n} = 
1$                \\
\hline
\end{tabular}
\caption{Different possible values for the parameters and the 
constraints imposed by \eqref{eq:constraint1}.
\label{tab:cons1}}
\end{centering}
\end{table}

We then obtain the following values of the coefficients:
\begin{align}
C_1^{(m)} &=  \frac{u_2}{v_1-u_2} P_{\alpha,\gamma}(e^s/u_{m+1}) \,,\\
D_1^{(m-1)} &= -\frac{v_1}{v_1-u_2}P_{\alpha,\gamma}^*(e^s/v_{m})\,,\\
\Lambda_1^{(m)} &= 
\frac{(q/p)^m(\Lambda_1+\alpha)v_1-(p/q)^m(\Lambda_1+\gamma)u_2
+\gamma u_2-\alpha v_1}{v_1-u_2}\,, \\
\widetilde{\Lambda}_1^{(m-1)} &=  \frac{(p/q)^m (\Lambda_1+\gamma)u_2 
- (q/p)^m v_1 (\Lambda_1+\alpha) - \gamma v_1 + 
(\alpha+q-p)u_2}{v_1-u_2}\,.
\end{align}

We also remind that for ASEP probabilistic models, all parameters 
have to be 
positive, so that only the two first lines of table \ref{tab:cons1} 
have to be 
considered in this case. These two constraints can be recasted into a 
single one
\beq
\label{eq:firstcond}
\Big(\frac{\alpha\beta}{\gamma\delta}e^s
-\left(\frac{q}{p}\right)^{L-1-n}\Big)
\Big(e^s-\left(\frac{q}{p}\right)^{L-1+n}\Big)=0\,.
\eeq
Then, using the correspondence (\ref{eq:asepXXZ}), it takes the form 
for XXZ model for even $L$:
\bea
&&\cos(\theta_{1}-\theta_{2})=\cos(\omega_{+}+\omega_{-}-m\,\eta)\,,
\eea
that is just the constraint given in \cite{nepo,Cao,dGE} for some integer $m$. 
Doing the same with the last two constraints, we get
\beq
\cos(\theta_{1}-\theta_{2})=\cos(\omega_{+}-\omega_{-}-n\,\eta)\,.
\eeq

\subsection{Second choice}
\label{sec:second}

Using the same technique as in the previous subsection, we introduce 
a basis of states that is suitable to the diagonalization of the 
alternative representation \eqref{eq:secondrepres} of the matrix $W$.
 The vector 
$\ket{V}_i$ has the same definition as before but we introduce the 
family of vectors
\begin{equation}
\label{eq:def:omegastar}
\ket{\omega^*(u)}_i = \begin{pmatrix} 1 \\ u \end{pmatrix}\,.
\end{equation}
We define similarly the product state $\ket{\Omega^*(u)}_i^j$ as in 
\eqref{eq:def:Omega} by replacing $\ket{\omega(u)}$ by 
$\ket{\omega^*(u)}$ and we introduce the new state with the new 
excitations in ordered positions $1\leq y_{m+1} < \ldots < y_n \leq 
n$:
\begin{equation}
\label{eq:def:state:star}
\ket{y_{m+1},\ldots,y_n}^* = 
\left(\sqrt{\frac{p}{q}}\right)^{(y_{m+1}-1)+\ldots+(y_n-1)} 
\ket{\Omega^*(u_{m+1}^*)}_1^{y_{m+1}-1}\ket{\omega^*(v_{m+1}^*)}_{y_{m+1}} 
\ldots \ket{\omega^*(v_n^*)}_{y_n}\ket{\Omega^*(u_{n+1}^*)}_{y_n+1}^L
\end{equation}
where the new coefficients $u_i^*$ and $v_i^*$ satisfy the recursion
relation
\begin{equation}
\label{eq:recursion:ui:vi:star}
u_{m+1}^* = (p/q) u_m^*\,, \quad v_{m+1}^* = (p/q) v_m^*\,.
\end{equation}

These vectors have been chosen such that we get for the Hamiltonian 
bulk part
\begin{equation}
\begin{split}
 w_{\beta,\beta+1}^*|y_{m+1},..,y_n\rangle^*=
\begin{cases}
0
 &\text{if } \beta,\beta+1\neq y_{m+1},\dots,y_n\,;\\
\sqrt{pq}|..y_j-1..\rangle^*
-p|..y_j..\rangle^* -|..\overline{y_j}..\rangle^*
&\text{if } \beta+1=y_j\ ;\,\beta\neq y_{j-1}\,;\\
\sqrt{pq}|..y_j+1..\rangle^*
-q|..,y_j,..\rangle^* +|..,\overline{y_j},..\rangle^*
&\text{if } \beta=y_j\ ;\,\beta+1\neq y_{j+1}\,;\\
 |..,\overline \beta,..\rangle^*
-|..,\overline{\beta+1},..\rangle^*
&\text{if } \beta=y_j\ ;\,\beta+1=y_{j+1}\,.
\end{cases}
\end{split}
\end{equation}

As for the previous choice, there are additional states $\ket{V}_1$ 
and $\ket{V}_L$ that survive the telescopic sum $\sum_{i=1}^{L-1} 
w_{i,i+1}^*$ and that must be absorbed by the boundary operators. We 
impose the dynamics:
\begin{align}
\label{eq:K1-2}
 &\widehat{K}_1^*|x_{m+1},\ldots\rangle^* =
\begin{cases}
 \Lambda_1^{*,(m)}|x_{m+1},\ldots\rangle^* 
+C_1^{*,(m)}|1,x_{m+1},\ldots\rangle^*  
&\text{if } x_{m+1}>1\,,\\
\widetilde{\Lambda}_1^{*,(m)}|1,x_{m+2}\dots\rangle^* 
+D_1^{*,(m)}
|x_{m+2},\ldots\rangle^*-|\overline{1},x_{m+2},\dots\rangle^*  
&\text{if } x_{m+1}=1\,,
\end{cases} \\
 &K_L^*|x_{m+1},\dots,x_n\rangle^* =
\begin{cases}
 \Lambda_L^*|x_{m+1},\dots,x_n\rangle^* &\text{if } x_{n}<L\,,\\
\widetilde\Lambda_L^*|x_{m+1},\dots,x_{n-1}, L\rangle^*
+|x_{m+1},\dots,x_{n-1},\overline 
L\rangle^* &\text{if } x_{n}=L \,,
\end{cases}\label{eq:KL-2}
\end{align}
with the additional closure constraint
\begin{equation}
\label{eq:closurestar}
C_1^{*,(0)}=0\,.
\end{equation}
We introduce again the more compact form 
$\Lambda_1^*=\Lambda_1^{*,(0)}$ 
since it is this value that will appear in the energy. {{From}} the 
constraints \eqref{eq:KL-2} and \eqref{eq:closurestar} and the 
notations introduced in section \ref{sec:boundarynotations}, we 
identify directly the value of most parameters:
\begin{align}
u_1^* &= e^s/c^*_\epsilon(\alpha,\gamma) \,, &\Lambda_1^* &= 
\lambda^*_\epsilon(\alpha,\gamma) \,,\\
u_{n+1}^* &= c^*_{\epsilon'}(\beta,\delta)  , & \Lambda_L^* &= 
\lambda^*_{\epsilon'}(\beta,\delta)+p+q \\
v_n^* &= c_{\epsilon''}(\beta,\delta)   \,, & 
\widetilde{\Lambda}_L^*&= p+\lambda_{\epsilon''}(\beta,\delta) 
\,.
\end{align}
As in the previous subsection, these values are compatible with the 
recursion relation \eqref{eq:recursion:ui:vi:star} if and only if the 
parameters of the model satisfy the relation
\begin{equation}
\label{eq:cont-2}
c^*_\epsilon(\alpha,\gamma)c^*_{\epsilon'}(\beta,\delta) = 
e^s\left(\frac{p}{q}\right)^{n}\,,
\end{equation}
for some signs $\epsilon$ and $\epsilon'$ and some integer $n$, which 
then fixes the number of excitations. Similar exceptional points 
already appeared in the literature in the context of the matrix 
ansatz described in section \ref{sec:matrixansatz}. Table 
\ref{tab:cons2} summarizes the different possibilities.

\begin{table}[h b t]
\begin{centering}
\begin{tabular}{|c  || c | c || c |}
 \hline
$\Lambda_1^*+\Lambda_L^*$  & $\epsilon$ &$\epsilon'$
& Constraints 
\\  
\hline
$ \frac{-\alpha-\gamma-\beta-\delta 
+\sqrt{(p-q+\gamma-\alpha)^2+4\gamma\alpha} + 
\sqrt{(p-q+\delta-\beta)^2+4\beta\delta}}{2}$  & $+$  & 
$+$ & 
$ c^*_+(\alpha,\gamma)c^*_+(\beta,\delta)
=e^s\left(\frac{p}{q}\right)^{n}$
\\  
\hline
$ \frac{-\alpha-\gamma-\beta-\delta 
-\sqrt{(p-q+\gamma-\alpha)^2+4\gamma\alpha} - 
\sqrt{(p-q+\delta-\beta)^2+4\beta\delta}}{2}$  & $-$  & 
$-$ & 
$ c^*_-(\alpha,\gamma)c^*_-(\beta,\delta)
=e^s\left(\frac{p}{q}\right)^{n}$\\  
\hline
$ \frac{-\alpha-\gamma-\beta-\delta 
+\sqrt{(p-q+\gamma-\alpha)^2+4\gamma\alpha} - 
\sqrt{(p-q+\delta-\beta)^2+4\beta\delta}}{2}$   & $+$  & 
$-$ & 
$c^*_+(\alpha,\gamma)c^*_-(\beta,\delta)
=e^s\left(\frac{p}{q}\right)^{n}$\\  
\hline
$\frac{-\alpha-\gamma-\beta-\delta 
-\sqrt{(p-q+\gamma-\alpha)^2+4\gamma\alpha} + 
\sqrt{(p-q+\delta-\beta)^2+4\beta\delta}}{2}$   & $-$  & 
$+$ & 
$c^*_-(\alpha,\gamma)c^*_+(\beta,\delta)
=e^s\left(\frac{p}{q}\right)^{n}$\\  
\hline
\end{tabular}
\caption{Different possible values for the parameters and the 
constraints imposed by \eqref{eq:constraint1}.
\label{tab:cons2}}
\end{centering}
\end{table}
The values of all the coefficients are then given by:
\begin{align}
C_1^{*,(m)} &= 
u^*_{2}\frac{P^*_{\alpha,\gamma}(e^s/u^*_{m+1})}{v^*_{1}-u^*_{2}} 
\,,\\
D_1^{*,(m-1)} &= 
-v^*_{1}\frac{P_{\alpha,\gamma}(e^s/v^*_{m})}{v^*_{1}-u^*_{2}}
\,,\\
\Lambda_1^{*,(m)} &= 
\frac{(p/q)^m v_1^* 
(\Lambda_1^*+\alpha+q)-(q/p)^mu_2^*(\Lambda_1^*+\gamma+p) + 
(\gamma+p)u_2^* - (\alpha+q) v_1^*}{v_1^* -u_2^*}
\,, \\
\widetilde{\Lambda}_1^{*,(m-1)} &= 
\frac{(q/p)^m u_2^� (\Lambda_1^*+\gamma+p) - (p/q)^m 
(\Lambda_1^*+\alpha+q)v_1^* + (\alpha+p)u_2^* - 
(\gamma+p)v_1^*}{v_1^*-u_2^*}
\,.
\end{align}

In the language of the XXZ spin chain, we should introduce the 
additional angles $\Omega_\pm$ defined by
\begin{equation}
2 \cos \Omega_\pm = \cos \delta_\pm - \cos \omega_\pm \,,
\end{equation}
and the condition \eqref{eq:cont-2} takes the following simple form 
for even $L$:
\begin{equation}
\cos(\theta_1-\theta_2) = \cos( \Omega_- \pm \Omega_+ -\eta m)\,,
\end{equation}
for some even integer $m$ between $-L$ and $L$. Up to our knowledge, 
this condition is new. The associated Bethe equations are presented 
in section \ref{subsec:betheeqs:secondpoints}.

\section{Coordinate Bethe Ansatz\label{sec:CBA}}

\subsection{The first set of specific points}
We are now in position to propose an ansatz for the eigenfunction
\begin{equation}
\label{eq:ansatz}
 \Phi_n=\sum_{m=0}^n\ \sum_{x_{m+1}<\dots<x_n}\ \sum_{g\in G_m}\
A_g^{(m)}\ e^{i\boldsymbol{k}^{(m)}_g.\boldsymbol{x}^{(m)}}\ 
|x_{m+1},\dots,x_n\rangle\,,
\end{equation}
where $G_m$ is a full set of representatives of the following coset 
$BC_n/BC_m$ (see appendix \ref{sec:BC}),
the vectors $|x_{m+1},\dots,x_n\rangle$ are either given by 
(\ref{eq:def:state}) or by (\ref{eq:def:state:star})
and we introduce the notation $\boldsymbol{k}^{(m)}$ for the 
following truncated vector
\begin{equation}
\boldsymbol{k}^{(m)}=(k_{m+1},\dots,k_n)\;.
\end{equation}
For this definition to be consistent, the coefficients $A_{g}^{(m)}$ 
do not have to depend on the choice of the representative i.e.
\begin{equation}
 A_{gh}^{(m)}=A_{g}^{(m)}\mb{for any $h\in BC_m$.}
\end{equation}

The coefficients $A_{g}^{(m)}$ are complex numbers to be determined 
such that $\Phi_n$ is
an eigenfunction of $H$ i.e. such that the following equation holds
\begin{equation}\label{eq:sch}
 H \Phi_n = E \Phi_n\;.
\end{equation}
Due to the results of section \ref{sec:basis}, $H\cB_n\subset \cB_n$.
Then, we project equation (\ref{eq:sch}) on the different independent 
vectors of $\cB_n$ to get constraints on the coefficients 
$A^{(m)}_{g}$.
This projection depends on the two choices done for the vectors 
$|x_{m+1},\dots,x_n\rangle$.
However, we chose the coefficients in such a way that one can be 
deduced from the other easily by adding a star to most quantities.
Therefore,
we write only the projections for the choice of section 
\ref{sec:first} and only the ones 
leading to independent relations (one can check that the remaining 
ones do not lead to new relations).

\paragraph{On $\boldsymbol{|x_{1},\dots,x_n\rangle}$ for 
$\boldsymbol{(x_{1},\dots,x_n)}$ generic} (i.e. $1<x_1$, $x_n<L$ and 
$1+x_j<x_{j+1}$)\\

As in the usual coordinate Bethe ansatz \cite{bethe}, this projection 
provides the energy:
\begin{equation} 
E=\Lambda_1+\Lambda_L+
\sum_{j=1}^n\lambda(e^{ik_j})
\mb{where}
\lambda(x)=\sqrt{pq}\big(x+\frac{1}{x}\big)-p-q
=\frac{\sqrt{pq}}{x}\left(x-\sqrt{\frac pq}\right)\left(x-\sqrt{\frac 
qp}\right)\,.
\label{def:lambdax}
\end{equation}
Let us remark that, up to the boundary terms $\Lambda_1$ and 
$\Lambda_L$, the energy takes 
the same form  as in the periodic case.

\paragraph{On $\boldsymbol{|x_{1},\dots,x_n\rangle}$ with 
$\boldsymbol{x_{j+1}=1+x_{j}}$} (and 
$x_1,\dots,x_{j-1},x_{j+2},\dots,x_n$ generic)\\

This projection is also a usual one and provides the scattering 
matrix between excitations. It is given by a relation between 
$A^{(0)}_{g}$ and $A^{(0)}_{g\sigma_j}$ where $\sigma_j$ is the 
permutation of $j$ and $j+1$ (see appendix \ref{sec:BC}).
Namely, we get
\begin{equation} 
\label{eq:S}
A^{(0)}_{g\sigma_j}
=S\left(e^{ik_{gj}},e^{ik_{g(j+1)}}\right)~A^{(0)}_{g}\,,
\end{equation}
with
\begin{equation}
 S(u,v)=-\frac{a(u,v)}
{a(v,u)}
\mb{where}
a(x,y)=\frac{i}{xy-1}\left(\left(\sqrt{\frac{q}{p}}
+\sqrt{\frac{p}{q}}\right)y-xy-1\right)\,.
\label{def:alphax}
\end{equation}
The normalisation chosen for the function $a(x,y)$ is for 
further simplifications.
As expected, this relation is similar to the periodic case since the 
boundaries 
are not involved in this process.

\paragraph{On $\boldsymbol{|1,x_{m+1}\dots,x_n\rangle}$} 
($x_{m+1},\dots,x_n$ generic and $m\geq1$)\\

This relation is a new one and we must take into account that the 
left 
boundary can create one particle
at the site $1$. We finally get, for any $g\in G_m$,
\begin{eqnarray}\label{eq:1g}
&&\Big(\widetilde{\Lambda}_1^{(m-1)}-\Lambda_1-p
+\sum_{j=1}^{m}\big(p+q-\sqrt{pq}(e^{ik_{gj}}+e^{-ik_{gj}})\big)\Big)
\sum_{h\in H_m} A_{gh}^{(m-1)}e^{ik_{ghm}}
\nonu
&&+\sqrt{pq}\sum_{h\in H_m}A_{gh}^{(m-1)}e^{2ik_{ghm}}
+C_1^{(m)}A_g^{(m)}=0\,,
\end{eqnarray}
where $H_m=BC_m/BC_{m-1}$.
To obtain this relation, we have used the following property 
\begin{equation}
 \sum_{g\in G_{m-1}}A_g 
e^{ik_{g(m)}}e^{i\boldsymbol{k}_{g}^{(m)}\boldsymbol{x}^{(m)}}
=\sum_{g\in G_{m}}e^{i\boldsymbol{k}_{g}^{(m)}\boldsymbol{x}^{(m)}}
\sum_{h\in H_m}A_{gh}e^{ik_{gh(m)}}\;.
\end{equation}
Let us stress that equation (\ref{eq:1g}) is invariant by the choice 
of representative of $G_m$.

\paragraph{On $\boldsymbol{|x_{m+1}\dots,x_n\rangle}$} 
($x_{m+1},\dots,x_n$ generic and $m\geq 1$)\\

This projection provides a second relation between the coefficient 
from the level $m-1$ and $m$ 
since we must take into account that the left boundary can destroy a 
particle present  on the site 1.
We obtain the following constraint, for any $g\in G_m$,
\begin{equation}\label{eq:2g}
 D_1^{(m-1)}\sum_{h\in H_m}A_{gh}^{(m-1)}e^{ik_{gh(m)}}
+(\Lambda_1^{(m)}-\Lambda_1+
\sum_{j=1}^{m}(p+q-\sqrt{pq}(e^{ik_{gj}}+e^{-ik_{gj}})))A_g^{(m)}=0
\,.
\end{equation}
{From} (\ref{eq:1g}) and (\ref{eq:2g}), we may express all the 
$A_g^{(m)}$ ($m\geq 1$) in terms of $A_g^{(0)}$
thanks to the following recursive relations defined for $m\geq 1$
\begin{equation}
\label{eq:recuT}
 A_g^{(m)}=T^{(m)}(e^{ik_{g1}},\dots,e^{ik_{gm}})A_g^{(m-1)}\,,
\end{equation}
with the following definitions:
\begin{align}
T^{(m)}(x_{1},\dots,x_{m}) &=\frac{D_1^{(m-1)}}{p_1(x_{m}) 
V_1(x_{m})}
\frac{x_m^2-1}{\prod_{j=1}^{m-1}a(x_{m},x_j)a(x_{j},1/x_m)} \,,\\
V_1(Z) &=\lambda(Z)
+ (\Lambda_1+\gamma) \Big(1-\frac{1}{Z}\sqrt{\frac{p}{q}}\Big)
+ (\Lambda_1+\alpha)\Big(1-\frac{1}{Z}\sqrt{\frac{q}{p}}\Big)\,,
\label{def:V1x} \\
p_1(Z) &=Z+ r \label{def:p1x}\,, \\
r &= \frac{1}{\sqrt{pq}} \frac{pu_2-q v_1}{v_1-u_2} = 
\frac{1}{\sqrt{pq}} \frac{pu_{m+1}-q v_m}{v_m-u_{m+1}} \,,
\label{def:coeff:r}
\end{align}
where $r$ describes the difference between $\ket{\omega(v_k)}$ and 
the 
vacuum states $\ket{\omega(u_k)}$ and $\ket{\omega(u_{k+1})}$. It is 
independent from $k$.

The proof that (\ref{eq:recuT}) is a solution of 
both equations (\ref{eq:1g}) and (\ref{eq:2g}) is postponed to  
appendix \ref{sec:B} and relies on a residue computation.
The integrability of the model plays a role at this place, since there 
are a priori too many constraints but not all of them are independent.

Let us emphasize that this recurrence relation (\ref{eq:recuT}) is 
the main result of this article.
Indeed, firstly, it proves that the ansatz 
(\ref{eq:ansatz}) and, in particular, the choice of cosets used to 
define it, is the appropriate choice. Secondly, it gives a very 
simple way to construct eigenfunctions, and we believe that it may be 
used for further computations.

A consequence of (\ref{eq:recuT}) (for $m=1$) and 
$A_{gr_1}^{(1)}=A_g^{(1)}$, is 
\begin{equation}
 A_{gr_1}^{(0)}
=\frac{T^{(1)}(e^{ik_{g1}})}{T^{(1)}(e^{-ik_{g1}})}A_g^{(0)}\,.
\end{equation}
This relation, together with (\ref{eq:S}), allow us to express $A_g^{(0)}$ for 
any $g\in BC_n$ in terms of $A_{1}^{(0)}$ 
(where the subscript $1$ stands for the unit of $BC_n$ group).
Finally, using recursively (\ref{eq:recuT}), we can express all the 
coefficients $A_g^{(m)}$ in terms of only $A_1^{(0)}$. This last 
coefficient is 
usually chosen such that the eigenfunction $\Phi_n$ be normed.

\paragraph{On $\boldsymbol{|x_{1}\dots,x_{n-1},L\rangle}$} 
($x_{1}\dots,x_{n-1}$ generic)\\

This last constraint consists in the quantization of the excitations moments 
since the 
system is in a finite volume. In the context of the coordinate Bethe 
ansatz,
this quantization leads to the so-called Bethe equations, explicitly 
given by, for $1\leq j \leq n$,

\begin{equation}
\label{eq:bethe}
 \prod_{\substack{\ell=1 \\ \ell\neq j}}^n 
S(e^{ik_\ell},e^{ik_j})S(e^{-ik_j},e^{ik_\ell})
=e^{2iLk_j}
\frac{V_1(e^{ik_{j}})V_L(e^{ik_{j}})}
{V_1(e^{-ik_{j}})V_L(e^{-ik_{j}})}\,,
\end{equation}
\begin{equation}
V_{L}(x) = \lambda(x)+
\left(\Lambda_{L}+\beta\right)
\Big(1-\frac 1x\sqrt{\frac qp}\Big)
+(\Lambda_{L}+\delta)\Big(1-\frac 1x\sqrt{\frac 
pq}\Big)\,.
\label{def:VLx}
\end{equation}
We remind that the parameters have to obey one of the relations given 
in table \ref{tab:cons1}.

\subsection{The new second set of specific points}
\label{subsec:betheeqs:secondpoints}
When condition \eqref{eq:cont-2} is satisfied, the same construction 
as in the previous section works, up to easy modifications in the 
equations.
We introduce  a new function $V_1^*(Z)$ and a new parameter $r^*$, 
which replaces $r$, defined by:
\begin{align}
\label{def:V1star}
V_1^*(Z) &= \lambda(Z) +(\Lambda_1^*+\alpha+q) \left( 
1-\frac{1}{Z}\sqrt{\frac{p}{q}}\right) + (\Lambda_1^*+\gamma+p) 
\left(1-\frac{1}{Z}\sqrt{\frac{q}{p}}\right) \,,\\
V_L^*(Z) &= \lambda(Z) +(\Lambda_L^*+\alpha+q) \left( 
1-\frac{1}{Z}\sqrt{\frac{p}{q}}\right) + (\Lambda_L^*+\gamma+p) 
\left(1-\frac{1}{Z}\sqrt{\frac{q}{p}}\right) \,,\\
r^* &= \frac{1}{\sqrt{pq}}\frac{pv_1^*-qu_2^*}{u_2^*-v_1^*}\,.
\end{align}
The same proof as the one presented in appendix \ref{sec:BC} is valid 
and we obtain the Bethe equations for $j\in\{1,\ldots,n\}$:
\begin{equation}
e^{2iL k_j} \frac{V_1^*(e^{ik_j})V_L^*(e^{ik_j}) }{V_1^*(e^{-ik_j})  
V_L^*(e^{-ik_j})} = \prod_{\substack{l=1 \\ l\neq j}}^n 
S(e^{ik_l},e^{ik_j})S(e^{-ik_j},e^{ik_l})\,,
\end{equation}
with the parameters obeying now one of the relations described in 
table \ref{tab:cons2}.

\section{Conclusions}

We discuss in this section two open problems, which need further investigation.

\subsection{Connection with the matrix ansatz}
\label{sec:matrixansatz}
For $s=0$, the matrix $W$ is stochastic (Markov transition matrix of 
the exclusion process) and hence has a known ground state eigenvalue 
$E=0$ with a simple left eigenvector, whose components are all equal 
to $1$ (conservation of total probability). Nevertheless, the 
corresponding right eigenvector is non-trivial since the matrix is 
not Hermitian and describes the stationary properties of the 
asymmetric exclusion process. The structure of this specific 
eigenvector was elucidated first in \cite{dehp} and then studied 
algebraically in more details in \cite{esr,mallicksandow}.

As already explained, the state space is $2^L$ dimensional and the 
canonical basis can be indexed by the values of the occupation 
$\tau_i\in\{0,1\}$ (resp. spin $s_i\in\{-1,1\}$ in the XXZ language) 
on each site $i$. The matrix ansatz states that the ground state of 
the ASEP with $s=0$ has components given by:
\begin{equation}
\langle \tau_1\tau_2\ldots\tau_L | \Phi \rangle = \langle\langle V_1 
| \prod_{1\leq i \leq n}^{\longrightarrow}\left( \tau_i D+ 
(1-\tau_i)E \right) | V_2 \rangle\rangle\,,
\end{equation}
where the arrow means that the product have to be build from left to 
right when the index $i$ increases. One has for example $\langle 
00\ldots00|\Phi\rangle= \langle\langle V_1 | E^L | V_2 
\rangle\rangle$. The non-commuting matrices $D$ and $E$ act on an 
abstract auxiliary vector space $\mathcal{V}$. The vector $| V_1 
\rangle\rangle$ is in this space $\mathcal{V}$, whereas the vector 
$\langle\langle V_2 |$ is in its dual. $\Phi$ is an eigenvector of 
$W$ for $E=0$ when $s=0$ if the two matrices $D$ et $E$ and the two 
boundary vectors satisfy the commutation rules \cite{dehp}:
\begin{subequations}
\label{matrixansatz}
\begin{align}
pDE-qED &= D+E \,,
\\
\langle\langle V_1 | (\gamma D - \alpha E) &= -\langle\langle V_1 |\,,
\\
(\beta D -\delta E) | V_2 \rangle\rangle &= | V_2 \rangle\rangle\,.
\end{align}
\end{subequations} with the condition $s=0$.
These three relations allow to determine recursively all the 
components of the eigenvector $\Phi$ and do not need an explicit 
representation of the algebra. In particular, the matrix ansatz gives 
an easy access to classical correlation functions with standard 
transfer matrix techniques.

The connection with the present Bethe ansatz approach comes from the 
algebraic study of the algebra generated by $D$ and $E$, as performed 
in \cite{esr,mallicksandow}, from computations that arise in the 
combinatorics of some so-called \emph{staircase tableaux} 
\cite{corteel}, as well as from the physical interpretation 
\cite{schuetz1,jafarpour,schuetz2} of the matrix ansatz in terms of 
so-called "shock" product state as introduced in section 
\ref{sec:basis}.

The first remark is that the matrix ansatz described here 
\emph{fails} for some values of the parameters 
$p,q,\alpha,\beta,\gamma,\delta$. Failure happens when the recursion 
relation
on the size $L$ induced by \eqref{matrixansatz} leads to 
$\langle\langle V_2 | V_1 \rangle\rangle=0$ and thus more generally 
to a null vector $\Phi$, as explained in \cite{esr}. This failure 
condition for a system of size $L$ is the existence of an integer 
$n\in\{0,1,\ldots, L-1\}$ such that 
\begin{equation}
\frac{\alpha\beta}{\gamma\delta} \left(\frac{p}{q}\right)^n = 1\,.
\end{equation}
This relation is precisely one of the two cases of 
\eqref{eq:firstcond} when $s=0$, for which the present Bethe ansatz 
approach works. 

The second remark is that the operators $D$ and $E$ do not have 
generically finite dimensional representations. Detailed studies 
shows that such finite-dimensional representations may exist at some 
specific points \cite{mallicksandow,esr}. It appears that an 
$n$-dimensional representation exists if one among the four cases of 
the condition \eqref{eq:cont-2} is satisfied for some integer $n$ and 
plus/minus signs, i.e. 
\begin{equation}
\label{eq:ma:finiterep}
c^*_+(\alpha,\gamma) c^*_+(\beta,\delta) = 
\left(\frac{p}{q}\right)^{n}.
\end{equation}
In \cite{mallicksandow}, the finite representation is used to study 
how the ground state evolves through the phase transition that occurs 
in the open ASEP along the manifold \eqref{eq:ma:finiterep}: the 
present paper gives also the excited states on the same manifold and 
the Bethe ansatz equations we obtain in this case now allow one to 
study how the gap behaves near this phase transition. Once again, 
atypical results appear in both the Bethe ansatz approach and the 
matrix ansatz approach for the same specific parameters.  

A third remark relies on the observation that the present 
construction does not allow to express the ground state described by 
the matrix ansatz in terms of a coordinate Bethe ansatz, precisely 
because one fails when the second works. This mismatch leads us to 
think that the matrix ansatz state may be used as a new reference 
state to build the missing eigenvectors if one could manage to add 
excitations on it. However, up to our knowledge, no deep 
understanding of the relation between these two approaches exists and 
has been exploited. 

A last remark deals with the combinatorics work \cite{corteel}. 
Besides the fact that quantities such as \eqref{eq:firstcond} and 
\eqref{eq:cont-2} appear in numerators or denominators in their 
computations and thus either simplifies or invalidates the matrix 
ansatz, their rewriting of the matrix ansatz in terms of 
\emph{staircase tableaux} sheds a new light on the precise structure 
of this ansatz, and puts it in a form that is more suitable to 
comparison with Bethe ansatz because of the relation of these 
tableaux with permutation tableaux. In particular, the number of 
so-called \emph{staircase tableaux} is $n!\,4^n$, whereas the cardinal 
of $BC_n$ is precisely $n!\,2^n$. Understanding both the similarities 
and the mismatch may lead to a better understanding of the Bethe 
ansatz.

\subsection{Completeness of the spectrum}
\label{subsec:completeness}
As usual for Bethe Ansatz methods, the delicate point to check is the completeness of the spectrum. Indeed, in many cases, completeness is not proved but expected to be true: numerics for small size systems \cite{wup}, enumeration of the number of roots in the thermodynamic limit \cite{KIRI}.

In the present case, numerical checks of the completeness for small sizes have been considered in \cite{completeness}: the full spectrum is shown to be described by two sets of Bethe equations. Our approach for the right eigenvectors gives only one set of Bethe equations for each specific point and thus only one part of the spectrum and one part of the right eigenvectors. The same construction can be done for left eigenvectors (or right eigenvectors of the adjoint). In this case, one obtains the other part of the spectrum and one part of the left eigenvectors, as discussed in \cite{simon09}.

Let us stress that the operators we are considering may not be hermitian and thus left and right eigenvectors may not be simply related. Although the full spectrum is known, it would be interesting to build the full set of right and left eigenvectors.

\section*{Acknowledgements}
This work was partially supported by the PEPS-PTI
grant \textit{Applications des Mod\`eles Int\'egrables}. We also 
thank Livia Ferro for a careful reading of the manuscript.

\appendix
\section{Weyl group $BC_n$ and coset\label{sec:BC}}

The Weyl group $BC_n$ is generated by the set 
$\{r_1,\sigma_1,\dots,\sigma_{n-1}\}$ with the following 
constraints:\begin{equation}
 \sigma_j^2=1=r_1^2\quad,\quad \sigma_1 r_1\sigma_1 
r_1=r_1\sigma_1r_1\sigma_1\quad,\quad
\sigma_j\sigma_{j+1}\sigma_j=\sigma_{j+1}\sigma_j\sigma_{j+1}\,.
\end{equation}
It acts on a vector $\boldsymbol{k}=(k_1,\dots,k_n)$ of 
$\mathbb{C}^n$:
\begin{equation}
 \boldsymbol{k}_{r_1}=(-k_1,k_2,\dots,k_n)\quad,\quad
 \boldsymbol{k}_{\sigma_j}=(k_1,\dots,k_{j+1},k_j,\dots,k_n)\,.
\end{equation}
The subgroup generated by $\{\sigma_1,\ldots,\sigma_{n-1}\}$ is just 
the symmetric group. We now consider its subgroups generated by 
$\{r_1,\sigma_1,\dots,\sigma_{m-1}\}$, $m\leq n$, 
which we identify with
$BC_m$.

For $g\in BC_n$, we then define the class $gBC_m=\{gh;\,h\in BC_m\}$, 
called a left coset. 
It is known that the set of all classes $gBC_n$, which is called 
$BC_n/BC_m$, forms a partition of
 $BC_n$: we can thus define $G_m$ as a full set of representatives of 
$BC_n/BC_m$, such that one 
has the unique decomposition $BC_n=\bigoplus_{g\in G_m} gBC_m$. We 
set, by convention, 
$G_0\sim BC_n$ and $G_n=\{1\}$.

The action of an element $g$ of $G_m$ on a vector
$\boldsymbol{k}^{(m)}=(k_{m+1},\dots,k_n)$ of $\mathbb{C}^{n-m}$
is given by $\boldsymbol{k}^{(m)}_g=(k_{g(m+1)},\dots,k_{g(n)})$. One 
checks that this action does not depend on the choice of the 
representative $g$, such that the action of $BC_n/BC_m$ is 
well-defined on $\mathbb{C}^{n-m}$ without further specifications.
This definition is useful because the set 
$\{\boldsymbol{k}^{(m)}_g|g\in G_m\}$ contains 
one and only once the vector 
$(\eps_{i_1}k_{i_1},\dots,\eps_{i_{n-m}}k_{i_{n-m}})$
for any choice $\eps_j=\pm$, $1\leq i_j\leq n$ and $i_j\neq i_k$.
For example, $\{\boldsymbol{k}^{(n-1)}_g|g\in 
G_{n-1}\}=\{(k_n),(-k_n),(k_{n-1}),(-k_{n-1}),\dots,
(k_1),(-k_1)\}$.

Finally, we introduce $H_m$ which is a full set of representatives 
of the coset $BC_m/BC_{m-1}$ used in section \ref{sec:CBA} and in the 
next appendix to simplify computations.

\section{Proof of solution (\ref{eq:recuT})\label{sec:B}}
We want to prove that (\ref{eq:recuT}) implies (\ref{eq:1g}) and 
(\ref{eq:2g}).\\

We start with (\ref{eq:2g}). We start by remarking that a consequence 
of 
(\ref{eq:recuT}) is
\begin{equation}
\label{eq:Sm}
 A_{g\sigma_\alpha}^{(m)}=
A_{g}^{(m)}\times\begin{cases}
 1 &1\leq \alpha \leq m-1\,,\\
\frac{T^{(m)}(e^{ik_{g1}},\dots,e^{ik_{gm-1}},e^{ik_{gm+1}})}
{T^{(m)}(e^{ik_{g1}},\dots,e^{ik_{gm}})}
S(e^{ik_{gm}},e^{ik_{gm+1}})& \alpha= m\,,\\
S(e^{ik_{g\alpha}},e^{ik_{g\alpha+1}})& \alpha\geq m+1\,.
\end{cases}
\end{equation}
Then, using again (\ref{eq:recuT}) to express now
$A_g^{(m+1)}$ in terms of $A_g^{(m)}$
and using (\ref{eq:Sm}) to express $A_{gh}^{(m)}$ in terms of 
$A_g^{(m)}$, relation 
(\ref{eq:2g}) becomes
the functional relation
\begin{eqnarray}\label{eq:ss1}
&&\sum_{j=1}^{m+1}\left[x_jV_1(x_j)p_1(x_j)
\prod_{\substack{\ell=1\\ \ell\neq 
j}}^{m+1}a(x_j,x_\ell)a(x_\ell,\frac{1}{x_j})
+\frac{1}{x_j}V_1(\frac{1}{x_j})p_1(\frac{1}{x_j})
\prod_{\substack{\ell=1\\ \ell\neq 
j}}^{m+1}a(\frac{1}{x_j},x_\ell)a(x_\ell,x_j)
-\lambda(x_j)\right]
\nonu
&&\qquad =\Lambda_1-\Lambda_1^{(m+1)}\,,
\end{eqnarray}
where $x_j$ stands for $\exp(ik_{gj})$ and the functions are defined 
in (\ref{def:alphax})-(\ref{def:V1x}).

To prove this last relation (\ref{eq:ss1}), let us introduce the 
following function
\begin{equation}
 F^{(m)}(Z)=\sqrt{pq}\frac{V_1(Z)p_1(Z)}
{\lambda(Z)\left(2Z-\sqrt{\frac{p}{q}}+\sqrt{\frac{q}{p}}\right)}
\prod_{\ell=1}^ {m+1} a(Z,x_\ell)a(x_\ell,\frac{1}{Z})\,,
\end{equation}
such that
\begin{eqnarray}
\res(F^{(m)}(Z))\Big|_{Z=x_j} &=& x_jV_1(x_j)p_1(x_j)
\prod_{\substack{\ell=1\\ \ell\neq 
j}}^{m+1}a(x_j,x_\ell)a(x_\ell,\frac{1}{x_j})\,,
\\
\res(F^{(m)}(Z))\Big|_{Z=1/x_j} &=& 
\frac{1}{x_j}V_1(\frac{1}{x_j})p_1(\frac{1}{x_j})
\prod_{\substack{\ell=1\\ \ell\neq 
j}}^{m+1}a(\frac{1}{x_j},x_\ell)a(x_\ell,x_j)\,,
\\
\res(F^{(m)}(Z))\Big|_{Z=\sqrt{\frac{p}{q}}}
&=& \sqrt{pq}
\left(\frac{q}{p}\right)^{m+1} \frac{\Lambda_1+\alpha}{p-q}\left( 
\sqrt{\frac{p}{q}}+r \right)\,,
\\
\res(F^{(m)}(Z))\Big|_{Z=\sqrt{\frac{q}{p}}}
&=& -\sqrt{pq}\left(\frac{p}{q}\right)^{m+1} 
\frac{\Lambda_1+\gamma}{p-q} \left( \sqrt{\frac{q}{p}}
+r\right)\,,
\\
\res(F^{(m)}(Z))
\Big|_{Z=\frac{p+q}{2\sqrt{{p}{q}}}}
&=&\frac{\sqrt{pq}}{2(p-q)} \left( p-q+2\gamma-2\alpha\right) 
\left( 
\frac{1}{2}\left(\sqrt{\frac{p}{q}}+\sqrt{\frac{q}{p}}\right)
+r\right)\,,
\\
\res(F^{(m)}(Z))\Big|_{Z=\infty}
&=& 
-\sum_{\ell=1}^{m+1}\lambda(x_\ell)
-\frac{\sqrt{pq}}{2}\left(\frac{1}{2}
\left(\sqrt{\frac{p}{q}}+\sqrt{\frac{q}{p}}\right)+r 
\right)-\frac{1}{2}(2\Lambda_1+\alpha+\gamma)\,.
\qquad
\end{eqnarray}
Then, (\ref{eq:ss1}) is equivalent to $\sum_{residue} F^{(m)}(x)=0$ 
which finishes the proof.

For (\ref{eq:1g}), we use the same procedure and (\ref{eq:ss1}) to get
\begin{eqnarray}\label{eq:ss2}
&&\sqrt{pq}\sum_{j=1}^{m+1}\left[x_j^2 V_1(x_j)p_1(x_j)
\prod_{\substack{\ell=1\\ \ell\neq 
j}}^{m+1}a(x_j,x_\ell)a(x_\ell,\frac{1}{x_j})
+\frac{1}{x_j^2}V_1(\frac{1}{x_j})p_1(\frac{1}{x_j})
\prod_{\substack{\ell=1\\ \ell\neq 
j}}^{m+1}a(\frac{1}{x_j},x_\ell)a(x_\ell,x_j)
\right]\nonumber\\
&&=\Big(\sum_{j=1}^{m+1}\lambda(x_j)+
\Lambda_1-\widetilde\Lambda_1^{(m)}+p\Big)
\Big(\sum_{j=1}^{m+1}\lambda(x_j)+\Lambda_1-\Lambda_1^{(m+1)}\Big)
-C_1^{(m+1)}D_1^{(m)}\,.
\end{eqnarray}
The function to consider is now 
$G^{(m)}(x)=\sqrt{pq}\,x\,F^{(m)}(x)$. Its 
residues at the point $x=x_{0}$ with $x_{0}\neq 0,\infty$ are simple 
to compute: $\res(G^{(m)}(x))\Big|_{x=x_{0}}=\sqrt{pq}\,x_{0}\, 
\res(F^{(m)}(x))\Big|_{x=x_{0}}$. Since 0 is not a pole, it remains 
to 
compute the residue at infinity. It reads:
\begin{eqnarray}
&& \res(G^{(m)}(x))\Big|_{x=\infty}
= \sum_{j, 
k=1}^{m+1} \lambda(x_{j})\,\lambda(x_{k}) + 2 
\left(\res(F^{(m)}(Z))\Big|_{Z=\infty}-\frac{p+q}{4} 
\right)\sum_{j=1}^{m+1}\lambda(x_{j}) \nonu
&& - 
\frac{\sqrt{pq}}{2}\left(r+\left(\sqrt{\frac{p}{q}}
+\sqrt{\frac{q}{p}}\right)\right)
\left(2\Lambda_1+\alpha+\gamma+\frac{p+q}{2}\right)
-\frac{(p-q)(\alpha-\gamma)}{4}+ 
\frac{(p+q)^2}{8}\,.\quad
\end{eqnarray}
Then, as above, (\ref{eq:ss2}) is equivalent to $\sum_{residue} G^{(m)}(x)=0$.

\end{document}